\documentclass{Generallatex}
\usepackage[numbers,sort&compress]{natbib}
\usepackage{graphicx}

\renewcommand{\ack}[1]{%
    \section*{Acknowledgments}
    #1
}

\begin{document}

\title{Enhancement of ${\bm J}_{\text{c}}$ by Proton Irradiation in HgBa\textsubscript{2}Ca\textsubscript{2}Cu\textsubscript{3}O\textsubscript{8+\bm{$\delta$}} Single Crystals}

\author{Wenjie Li$^{1,\ast}$\addorcid{0000-0001-8451-126X}{Wenjie Li}, Ran Guo$^{1}$, Xin Zhou$^{1}$, Qiang Hou$^{1}$, Mengqin Liu$^{1}$, Longfei Sun$^{1}$, Yuhang Zu$^{2}$, Wenshan Hong$^{3}$, \\Yuan Li$^{3}$, Sheng Li$^{1,4}$, Yue Sun$^{1,\dagger}$\addorcid{0000-0002-5189-5460}{Yue Sun}, Zhixiang Shi$^{1,\ddagger}$\addorcid{0000-0003-3881-5152}{Zhixiang Shi}, and Tsuyoshi Tamegai$^{2}$}

\affil{\small $^1$Key Laboratory of Quantum Materials and Devices of Ministry of Education, School of Physics, \\Southeast University, Nanjing 211189, China\\
$^2$Department of Applied Physics, The University of Tokyo, 7-3-1 Hongo, Bunkyo-ku, Tokyo 113-8656, Japan\\
$^3$Beijing National Laboratory for Condensed Matter Physics, Institute of Physics, \\Chinese Academy of Sciences, Beijing 100190, China\\
$^4$Purple Mountain Laboratories, Nanjing 211111, China}

\email{$\ast$ wenjiecd@gmail.com}
\email{$\dagger$ sunyue@seu.edu.cn}
\email{$\ddagger$ zxshi@seu.edu.cn}

\begin{abstract}
Critical current density is the key parameter for the practical application of superconductivity. In this study, 3 MeV proton irradiation experiments were conducted on HgBa$_2$Ca$_2$Cu$_3$O$_{8+\delta}$ single crystals to introduce pinning centers. The critical current density is found to be strongly enhanced after the irradiation with its maximum at a dose of $1\times10^{16}$/cm$^2$, where the self-field critical current density at 2 K is enhanced from 5.5 MA/cm$^2$ to 26 MA/cm$^2$. At $T = 77$ K, the self-field critical current density for all irradiated crystals is over 0.1 MA/cm$^2$. The power-law dependence of the critical current density on the magnetic field is observed after irradiation, with a large power-law exponent $\alpha$ close to 1. A monotonic magnetic field dependence of the normalized magnetic relaxation rate is observed, which could be attributed to the low irreversibility field caused by the large anisotropy in Hg1223 single crystals. Through the analysis of the pinning force density of the crystal before and after irradiation, a clear mechanism change has been observed.
\end{abstract}

\section{Introduction}
Cuprates superconductors have attracted significant research interest due to their high critical transition temperature $T_{\rm c}$, which exceeds the boiling point of nitrogen, demonstrating strong potential for practical applications. A major topic in the practical application of superconductivity has been the enhancement of the critical current density $J_{\rm c}$. The key factor limiting $J_{\rm c}$ in superconductors is the energy dissipation caused by the motion of magnetic flux quanta (vortices) formed under an applied magnetic field \cite{abrikosov2004nobel}. Suppressing vortex motion is commonly achieved by introducing pinning centers that locally reduce the free energy. Various types of defects can act as pinning centers, such as point, columnar, and planar defects \cite{kwok2016vortices,blatter1994vortices,Mchenry1994flux}. Point defects can be introduced via chemical doping \cite{Jinyulin2022strong} or light-particle irradiations (e.g., with electrons, protons, or neutrons) \cite{tamegai2012effects,cho2018using,Eisterer2002Neutron}. Columnar defects are typically created using heavy-ion irradiations \cite{Sun2017Columnar}. Planar defects may be formed by controlling crystal growth conditions such as annealing processes \cite{Ren2024Vortex}. Particle irradiation experiments have been conducted on various superconductors, and the introduced pinning centers have indeed led to notable enhancements in $J_{\rm c}$. For example, in Ref. \cite{civale1991vortex}, 3 MeV proton irradiation increased the self-field $J_{\rm c}$ of YBCO (YBa$_2$Cu$_3$O$_7$) single crystals at 5 K by approximately a factor of 5, from $\sim$1.6 MA/cm$^2$ to $\sim$8.4 MA/cm$^2$. A recent review paper related to the effects of particle irradiation on YBCO can be found in Ref. \cite{Civale2025sust}. The self-field $J_{\rm c}$ of MgB$_2$ single crystals is enhanced from 0.1 MA/cm$^2$ to 0.47 MA/cm$^2$ at 5 K after neutron irradition \cite{Eisterer2005MgB2}. The iron-based superconductors discovered in 2008 opened a new road for achieving practical applications of superconductivity \cite{hosono2018recent}. Many particle irradiation experiments have been conducted. For example, in Ref. \cite{tamegai2012effects}, 3 MeV proton irradiation enhanced the $J_{\rm c}$ of Ba(Fe$_{0.93}$Co$_{0.07}$)$_2$As$_2$ single crystals from 1 MA/cm$^2$ to 2.4 MA/cm$^2$ under 2 K at self-field at a dose of $1.2\times10^{16}$/cm$^2$. In Ref. \cite{Kihlstrom2013Bak1224MeV}, the self-field $J_{\rm c}$ of (Ba$_{0.6}$K$_{0.4}$)Fe$_2$As$_2$ single crystal is enhanced from 1.4 MA/cm$^2$ to 6.1 MA/cm$^2$ by 4 MeV proton irradiation at a dose of $7\times10^{16}$/cm$^2$. In FeSe single crystal, the self-field $J_{\rm c}$ at 2 K was enhanced from $\sim$0.03 MA/cm$^2$ to $\sim$0.08 MA/cm$^2$ by 3 MeV proton irradiation at a dose of $5\times10^{16}$/cm$^2$ \cite{Sun2015FeSealpha}. In FeSe$_{0.5}$Te$_{0.5}$ thin film, self-field $J_{\rm c}$ at 4.2 K is enhanced from $\sim$0.9 MA/cm$^2$ to $\sim$1.4 MA/cm$^2$ at a dose of $0.1\times10^{16}$/cm$^2$ \cite{Ozaki2016FeSeTeproton}. All these experiments show that introducing pinning centers through particle irradiations is an effective way to enhance $J_{\rm c}$ across diverse classes of superconductors.

Hg1223 (HgBa$_2$Ca$_2$Cu$_3$O$_{8+\delta}$) is the superconductor which holds the highest $T_{\rm c}$ under ambient pressure. It was discovered in 1993 \cite{Schilling1993Hg1223}, just a few years after the discovery of cuprate superconductor in 1986 \cite{Bednorz1986Possible}. Through the application of hydrostatic pressure, the $T_{\rm c}$ of Hg1223 polycrystal had been enhanced to over 160 K \cite{Gao1994Super}. However, due to the challenge of high Hg vapor during synthesis, obtaining pure Hg1223 single crystals directly is very difficult. Many researchers have attempted to substitute Hg with other elements to reduce the Hg vapor pressure during sample growth. There are several successful experiments such as through Pb doping \cite{Isawa1994Pb-doping}, Re doping \cite{Ueda2007Synthesis,Krelaus1999Magneti}, and Sb doping \cite{Li1998Enhancement}. For practical applications, significant progress has also been made in synthesizing thin films. For example, the self-field $J_{\rm c}$ of (Hg,Re)1223 thin films at 80 K exceeds 1 MA/cm$^2$, far surpassing the level for practical applications \cite{Moriwaki1999transport}. 

Recently, through improvements in the synthesis process, high-quality Hg1223 single crystals have been successfully synthesized \cite{Loret2017crystal,Wang2018Growth}. This breakthrough has reignited researchers' interest in studying Hg1223 superconductors \cite{Wen2025Unprecedentedly,Ye2025Three,oliviero2024charge,Mor2024Dome}. However, relatively little attention has been paid to the vortex dynamics in Hg1223 single crystals. So in this paper, we report the study on the mechanisms of vortex pinning and vortex dynamics in Hg1223 single crystals. By using 3 MeV proton irradiation, we successfully introduced artificial pinning centers in Hg1223 single crystals. These pinning centers are found to significantly enhance $J_{\rm c}$. 

\section{Experimental Details}
HgBa$_2$Ca$_2$Cu$_3$O$_{8+\delta}$ single crystals with $T_{\rm c}$ $\sim$129 K were synthesized by high-pressure reaction after preparing the precursors. The details of the crystal growth can be found in Ref. \cite{Wang2018Growth}. The as-grown crystals were subsequently annealed in flowing oxygen atmosphere at 500\,°C for 5 days to optimize the superconducting properties. It should be noted that in Hg1223, the inner and outer CuO$_2$ planes exhibit different hole doping levels due to their distinct local environments. A detailed determination of the doping distribution is beyond the scope of this work, but the bulk $T_{\text{c}}$ of $\sim$129\,K indicates that the overall doping is near optimal. The 3 MeV proton irradiation experiments were conducted at NIRS-HIMAC in Chiba, Japan. Before the irradiation, single crystals were prepared into thin plates with thickness of $\sim$20 $\mu$m, which is thinner than the projected range of 3 MeV protons in Hg1223 ($\sim$54 $\mu$m). The typical lateral dimensions of the samples are approximately $250\ \mathrm{\mu m} \times 150\ \mathrm{\mu m}$. An optical microscope image of a pristine Hg1223 single crystal is shown in the inset of Fig. \ref{fig:1}(a), revealing the typical morphology of the sample. The projected range is calculated by the stopping and range of ions in matter (SRIM-2008) \cite{ziegler1985stopping}. The magnetization measurements were performed with a SQUID magnetometer (MPMS, Quantum Design). $J_{\rm c}$ is estimated from the measured magnetic moment by using the extended Bean model~\cite{bean1964magnetization}. According to this model, $J_{\rm c}$ (A/cm$^{2}$) is given by:
\begin{equation}
J_{\text{c}} = \frac{20 \Delta M}{a \left(1 - \frac{a}{3b}\right)} \qquad (a < b),
\end{equation}
where $\Delta M$ (emu/cm$^{3}$) is the difference in magnetization when sweeping the external field down and up, with $a$ (cm) and $b$ (cm) are the width and length of the sample, respectively. $T_{\rm c}$ is defined as the onset temperature of the diamagnetic transition. For the magnetic relaxation measurements, temporal evolution of the magnetization was recorded for $\sim$1 hour. 

\section{Results and Discussion}
\begin{figure*}
\centering
\includegraphics[scale=0.9]{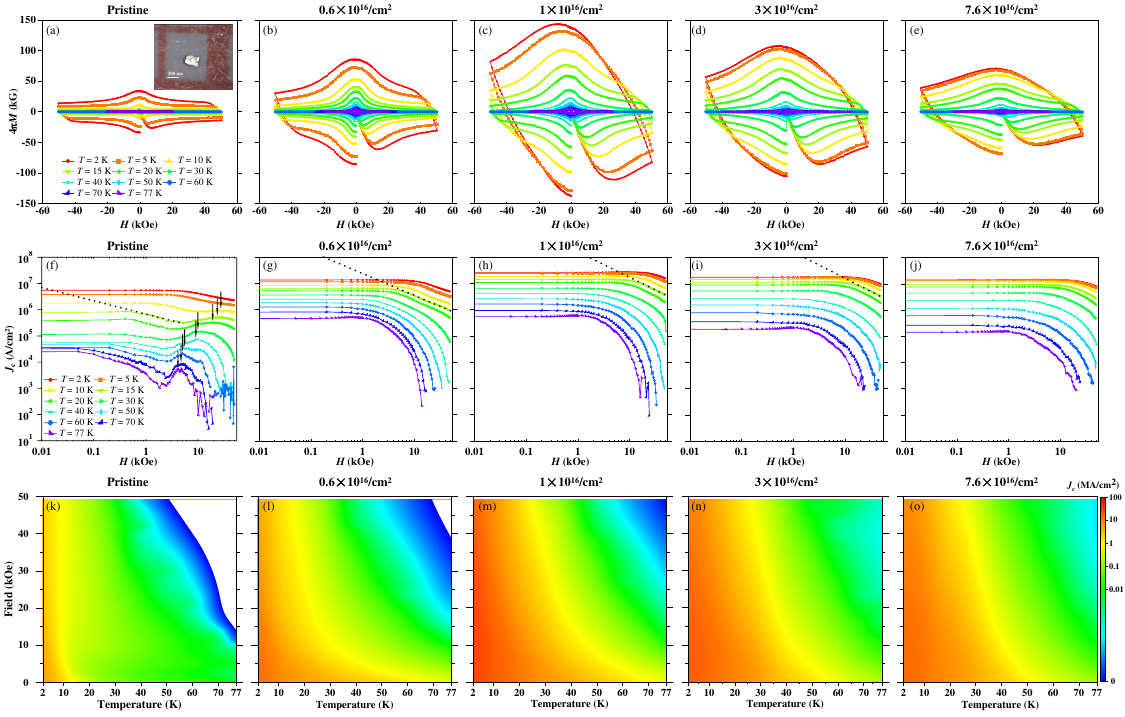}
\caption{\label{fig:1}(a)-(e) Isothermal external magnetic field dependence of magnetization and (f)-(j) $J_{\rm c}$ for Hg1223 pristine sample and samples irradiated by 3 MeV protons with dose of $0.6\times10^{16}$/cm$^2$, $1\times10^{16}$/cm$^2$, $3\times10^{16}$/cm$^2$, and $7.6\times10^{16}$/cm$^2$. (k)-(o) Contour map of $J_{\rm c}$ for Hg1223 single crystals before and after 3 MeV proton irradiation under different temperatures and magnetic fields. From these contour maps the irreversibility field $H_{\text{irr}}$ can be derived and a significant enhancement of $H_{\text{irr}}$ after proton irradiation is clearly observed. The inset in (a) is an optical microscope image of a pristine Hg1223 single crystal, showing the typical morphology of the sample. The sample dimensions are approximately $250\ \mathrm{\mu m} \times 150\ \mathrm{\mu m} \times 20\ \mathrm{\mu m}$. Black arrows in (f) indicate the positions of $J_{\rm c}$ peak at each temperature.} 
\end{figure*}

Figures \ref{fig:1}(a)-(j) show curves of isothermal magnetization and $J_{\rm c}$ of Hg1223 as a function of applied magnetic field in the range of 2–77 K before and after irradiation. Before the irradiation, the self-field $J_{\rm c}$ at 2 K, 5 K, and 77 K is 5.5 MA/cm$^2$, 3.8 MA/cm$^2$, and 0.02 MA/cm$^2$, respectively. The $J_{\rm c}$ in the pristine crystal is comparable to other cuprate superconductors. For example, the $J_{\rm c}$ for pure (undoped) Bi$_{2}$Sr$_{2}$Ca$_{2}$Cu$_3$O$_y$ single crystal is 0.3 MA/cm$^2$ at $T = 20$ K under $H = 5$ kOe \cite{shimoyama2002FluxPinningProperties}, while this value is 0.2 MA/cm$^2$ in pristine Hg1223. For NdBa$_2$Cu$_3$O$_{7-\delta}$ single crystal, the self-field $J_{\rm c}$ at 77 K is 0.03 MA/cm$^2$ \cite{Koblischka1998BSCOOThree}, which is close to that for Hg1223. The $J_{\rm c}$ we have discussed up to now is based on the condition where the vortices are completely depinned from the pinning centers, known as the depinning $J_{\rm c}$. On the other hand, there is another critical current density, called depairing critical current density $J_{\rm dp}$, where kinetic energy of Cooper pairs cancels the energy gain in the superconducting state. Here, we calculated the depairing $J_{\rm c}$ of the Hg1223 according to formula of
\begin{equation}
J_{\rm dp} = \frac{c\Phi_0}{12\sqrt{3}\pi^2\xi(0)\lambda(0)^2},
\end{equation}
where $c$ is the speed of light, $\Phi_0$ is the quantum of flux $\sim 2.07\times10^{-7}$ G$\cdot$cm$^2$, $\xi$ is a coherence length and $\lambda$ is a penetration depth. By using the data in Ref. \cite{Charles2000Handbook}, the calculated $J_{\rm dp}$ of Hg1223 is $\sim$460 MA/cm$^2$, which is much higher than our experimentally measured depinning $J_{\rm c}$. We also calculated the $J_{\rm dp}$ for several other typical superconductors using this formula, as listed in Table \ref{Table1}. Among these superconductors, Hg1223 exhibits the largest depairing $J_{\rm c}$. This suggests that the high depairing current density makes Hg1223 an attractive candidate for further $J_{\rm c}$ enhancement through the introduction of artificial pinning centers. 

\begin{table}[b]
    \centering
    \caption{Depairing critical current density $J_{\rm dp}$ with the $T_{\rm c}$, coherence length $\xi$ and penetration depth $\lambda$ for several kinds of superconductors. Reference data from \cite{Charles2000Handbook,Sun2020depairing,prozorov2009AnisotropicLondonPenetration,yuan2009NearlyIsotropicSuperconductivity,miura2024QuadruplingDepairingCurrent}. All parameters are cited from literature for optimally-doped or near-optimally-doped samples.} \label{Table1}
    \resizebox{0.48\textwidth}{!}{%
    \begin{tabular}{c|c|c|c|c}
    \hline
    Superconductor & $T_{\rm c}$(K) & $\xi$(nm) & $\lambda$(nm)  & $J_{\rm dp}$ (MA/cm$^2$) \\
    \hline
    HgBa$_2$Ca$_2$Cu$_3$O$_{8+x}$     & 133 & 1.3 & 130 & 460 \\
    \hline
    SmFeAsO$_{0.63}$H$_{0.37}$     & 46 & 1.5 & 125 & 430 \\
    \hline
    Nb$_3$Sn     & 16 & 2.8 & 93 & 420\\
    \hline
    YBa$_2$Cu$_3$O$_{8+x}$   & 91 & 1.65 & 156 & 250 \\
    \hline
    NbTi     & 9.6 & 3.8 & 130 & 160  \\
    \hline
    Bi$_2$Sr$_2$CaCu$_2$O$_x$ & 89 & 1.8 & 250 & 90  \\
    \hline
    Ba$_{0.6}$K$_{0.4}$Fe$_2$As$_2$   & 28 & 2.5 & 220 & 73 \\
    \hline
    FeSe     & 9 & 4.3 & 450 & 12  \\
    \hline
    \end{tabular}%
    }
\end{table}

\begin{figure}
\centering
\hspace{-7.5mm}\includegraphics[scale=0.7]{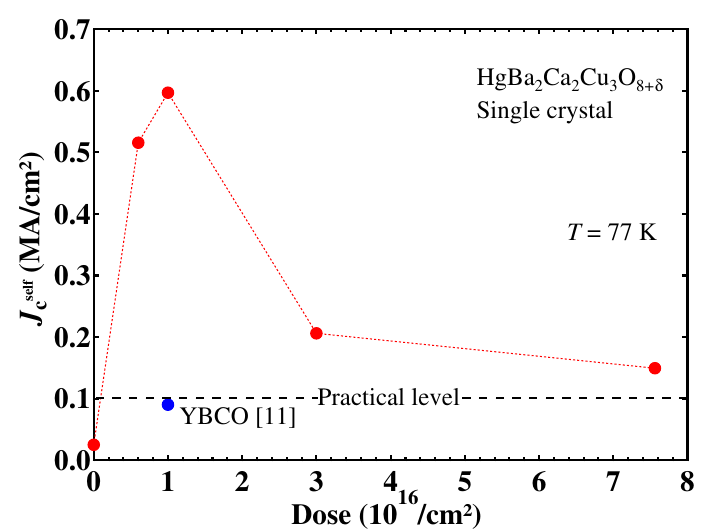}
\caption{\label{fig:2} Dose dependence of self-field $J_{\rm c}$ for Hg1223 single crystals at $T = 77$ K. The $J_{\rm c}$ of all irradiated samples exceeds the practical level. The self-field $J_{\rm c}$ at $T = 77$ K for YBCO single crystal after 3 MeV proton irradiation under the dose of $1\times10^{16}$/cm$^2$ is 0.09 MA/cm$^2$ \cite{civale1991vortex}.}
\end{figure}

Figures \ref{fig:1}(k)-(o) show the contour map of $J_{\rm c}$ in Hg1223 single crystals before and after 3 MeV proton irradiation under different temperatures and magnetic fields. From these figures, it is clear that $J_{\rm c}$ is significantly enhanced after 3 MeV proton irradiation. At 77 K, the self-field $J_{\rm c}$ already exceeds 0.5 MA/cm$^2$ under the dose of $1\times10^{16}$/cm$^2$, a value that even approaches the $J_{\rm c}$ of Hg1223 thin films under the same condition \cite{Gapud19971MeV}. A summary of $J_{\rm c}$ under the self-field is shown in Fig. \ref{fig:2}, where we can see that with increasing irradiation dose, $J_{\rm c}$ initially increases, reaching a maximum at a dose of $1\times10^{16}$/cm$^2$, and then begins to decline as the dose continues to increase. In Ref. \cite{civale1991vortex}, a systematic study on the effects of 3 MeV proton irradiation on YBCO single crystals was conducted, where the maximum $J_{\rm c}$ was found at a dose of $1\times10^{16}$/cm$^2$, and the self-field $J_{\rm c}$ at 77 K was 0.09 MA/cm$^2$ as shown in Fig. \ref{fig:2}. The $J_{\rm c}$ enhancement of Hg1223 single crystals after irradiation in this study shows larger values. The self-field $J_{\rm c}$ at 77 K for all irradiated Hg1223 single crystals exceeds the practical level of 0.1 MA/cm$^2$.

\begin{figure}
\centering
\hspace{-4mm}\includegraphics[scale=0.7]{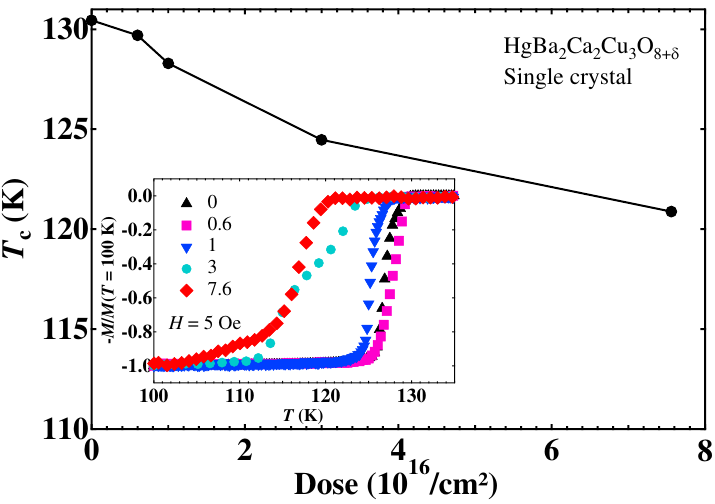}
\caption{\label{fig:3}Dose dependence of $T_{\rm c}$ for Hg1223 single crystals irradiated by 3 MeV protons. Inset is the raw $M$-$T$ curves measured under the external magnetic field of 5 Oe along sample $c$-axis. $T_{\rm c}$ is defined as the onset temperature of the diamagnetic transition.}
\end{figure}

\begin{figure*}
\centering
\includegraphics[scale=0.8]{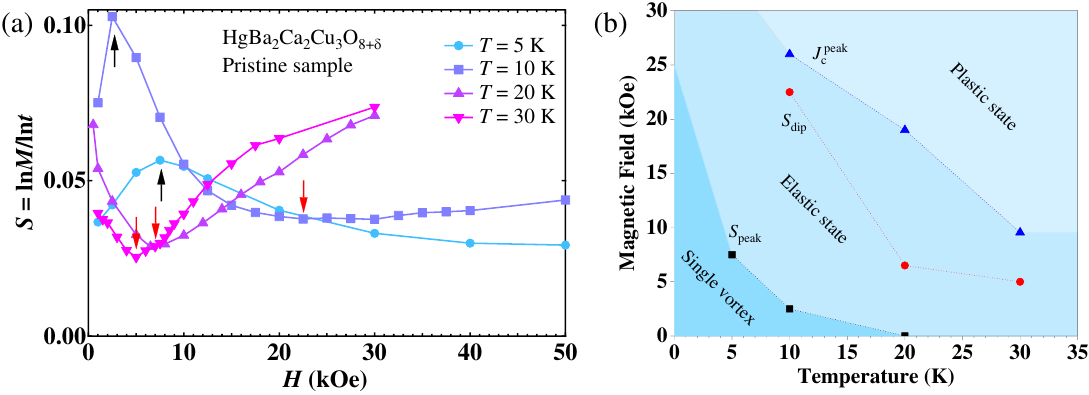}
\caption{\label{fig:4}(a) External magnetic field dependence of magnetic relaxation rate for Hg1223 single crystals before 3 MeV proton irradiation. Red arrows indicate the dip where the vortices are ideally straightened along the magnetic field direction, while black arrows indicate the peak where the vortex elastic state begins to form. (b) Schematic phase diagram for Hg1223 single crystal.}
\end{figure*}

The suppression trend of $J_{\rm c}$ at higher doses can be understood as the result of detrimental effect of irradiation on $T_{\rm c}$. Figure \ref{fig:3} depicts the variation of $T_{\rm c}$ with irradiation dose in Hg1223 single crystals, where $T_{\rm c}$ is gradually suppressed with increasing irradiation dose. $T_{\rm c}$ in superconductors is affected by various factors. One possible explanation is that introduced point defects can act as pair-breakers, weakening the $T_{\rm c}$ of the superconductor. In superconductors with nodes in the gap or anisotropic gaps, nonmagnetic disorder can suppress $T_{\rm c}$. Similar results of $T_{\rm c}$ suppression due to light-particle irradiation have been reported such as in FeSe \cite{Sun2017FeSeH+}, NbSe$_2$ \cite{li2023effects,cho2018using}, V$_3$Si \cite{Cho2022V3Si}, Ba$_{1-x}$K$_x$Fe$_2$As$_2$ \cite{Taen2013Pair}, and NdFeAs(O,F) \cite{Tarantini2010NdFeAs(OF)}. All these superconductors are either anisotropic superconductors and/or superconductors with multiple gaps. Another explanation is the presence of competing phases like CDW (charge-density wave) within the superconductor, which may compete with superconductivity for the Fermi surface. Generally, the introduction of point defects suppresses CDW more effectively, potentially freeing the gapped Fermi surface to form superconductivity. In Hg1223, there has been report of the existence of CDW. In Ref \cite{Loret2019HgCDW}, CDW was reported by measurement of Hg1223 through electronic Raman spectroscopy. It is possible that the introduced point defects have suppressed the CDW and released a portion of the Fermi surface, which in turn could enhance the $T_{\rm c}$ of Hg1223. Since the enhancement of $T_{\rm c}$ is less than the suppression caused by those pair-breakers, the overall trend of $T_{\rm c}$ appears as a monotonic decrease. While given the limited number of data points and the transition broadening at higher doses as well as the difficulty of performing normal-state resistivity measurements on such small crystals, we cannot draw a definitive conclusion about the suppression rate at low doses. A systematic resistivity study through light-particle irradiation experiments, similar to those reported in Ref. \cite{li2023effects,cho2018using,Sun2017FeSeH+,sun2025PbTaSe,nakajima2010SuppressionCriticalTemperature,Taen2013Pair,Tarantini2010NdFeAs(OF),Cho2022V3Si,park2018QuasiparticleScattering3} would be beneficial in elucidating the relationship between CDW and superconductivity in Hg1223. The inset in Fig. \ref{fig:3} presents the $M$-$T$ curves, showing that when the dose is less than $1\times10^{16}$/cm$^2$, the width of the superconducting transition remains nearly unchanged, but at doses of $3\times10^{16}$/cm$^2$ and $7.6\times10^{16}$/cm$^2$, the transition width significantly broadens, indicating that the irradiation damage to the superconducting sample becomes serious. Regarding the nature of the defects introduced by 3 MeV proton irradiation, previous systematic studies in other layered superconductors have been conducted. In BaFe$_2$(As$_{0.67}$P$_{0.33}$)$_2$ \cite{park2018QuasiparticleScattering3}, TEM imaging revealed black speckles in the irradiated sample that were not observed in the pristine sample, and these features correspond to nanometer-scale defect clusters with a maximum diameter up to 5 nm in the irradiated crystals. Geometrical phase analysis confirmed an increase in in-plane strain after irradiation. In 2$H$-NbSe$_2$ \cite{li2023effects}, a monotonic lattice expansion upon 3 MeV proton irradiation was observed. These observations suggest that 3 MeV proton irradiation mainly introduces nanometer-scale defect clusters rather than isolated point defects (Frenkel pairs). In the context of Hg1223, similar defect structures are expected. Transition broadening induced by such large scale defects have also been observed in other superconductors under particle irradiations \cite{li2022SuppressionSuperconductivityHeavy,park2018QuasiparticleScattering3,Sun2017Columnar}. This may explain the reduction in $J_{\text{c}}$ observed at doses larger than $1\times10^{16}$/cm$^2$.

From the $J_{\rm c}$-$H$ curves for pristine sample in Fig. \ref{fig:1}(f), the curves can be roughly divided into three regions based on the magnetic field. At low magnetic fields, the $J_{\rm c}$ remains almost constant. In this region, the interactions between vortices are negligible due to their large distance. In such a case, which is known as single-vortex pinning regime, each vortex is exclusively pinned by collective action of pinning centers. However, as the magnetic field increases, the repulsion between vortices becomes significant, which affects vortex motion as a whole. A notable feature in this regime is the frequent observation of a power-law dependence of the $J_{\rm c}$-$H$  \cite{Beek2002YBCOalpha}. The value of the power-law exponent $\alpha$ helps characterize the pinning feature. As indicated by the dashed line in the $T = 15$ K curve in Fig. \ref{fig:1}(f), where the $\alpha$ value is approximately -0.5. This value is similar to previous studies on YBCO thin films \cite{Beek2002YBCOalpha}, (Ba,K)Fe$_2$As$_2$ \cite{taen2015CriticalCurrentDensity} and Ba(Fe,Co)$_2$As$_2$ \cite{taen2012powerlaw} single crystals, which can be attributed to strong pinning caused by sparse nanometer-sized defects. With further increase in magnetic field, the pinning mechanism in the vortex system becomes more complex and the effect of the relaxation of vortices cannot be neglected. In this study, within the third region a clear peak effect in $J_{\rm c}(H)$ is observed for the pristine sample at several temperatures as indicated by the black arrows in Fig. \ref{fig:1}(f). The peak field $B_p$ decreases monotonically with increasing temperature. Following the analysis of the peak effect in YBa$_2$Cu$_3$O$_{7-x}$ \cite{abulafia1996PlasticVortexCreep}, the temperature dependence of $B_p$ can be well described by $B_p \propto [1-(T/T_{\rm c})^4]^2$, which is characteristic of the crossover from elastic to plastic vortex creep. This peak therefore marks the boundary between the elastic and plastic vortex regimes.

To understand the complex magnetic field dependence of $J_{\rm c}$, we performed magnetic relaxation measurements. Magnetic relaxation is primarily induced by thermal fluctuations and quantum tunneling \cite{yeshurun1996magnetic}. While vortices are pinned by pinning centers, parts of the vortices can move away from the pinning centers, leading to increased energy dissipation and a consequent reduction in $J_{\rm c}$. Therefore, it is generally observed that regions with higher relaxation rates tend to exhibit lower $J_{\rm c}$. To quantify this relaxation strength, the normalized relaxation rate $S$ is typically evaluated \cite{yeshurun1996magnetic}. The basic mechanism starts with the hopping time $t$ for a vortex to escape its pinning potential $U$ under thermal activation:
\begin{equation}
t = t_0 \exp\left(\frac{U}{kT}\right),
\end{equation}
where $t_0$ is the effective attempt time, $k$ is the Boltzmann constant, and $T$ is the temperature. Assuming the energy barrier $U$ decreases linearly with the flowing current $J$ with the form of
\begin{equation}
U = U_0 \left(1 - \frac{J}{J_{c0}}\right), 
\end{equation}
where $U_0$ is the barrier height at zero current and $J_{c0}$ is the critical current density at $t = 0$ s, combining these relations yields the classic Anderson-Kim equation for flux creep:
\begin{equation}
J = J_{c0} \left[1 - \frac{kT}{U_0} \ln\left(\frac{t}{t_0}\right)\right]. \label{Jc}
\end{equation}
From Eq. (\ref{Jc}), it is clear that the persistent current is decaying logarithmically with time. To remove the dependence on the unknown parameter $t_0$ and isolate the intrinsic pinning strength, the normalized relaxation rate $S$ is defined as the logarithmic derivative of the irreversible magnetization $M_{\text{irr}}$ (which is proportional to $J$):
\begin{equation}
S \equiv \left| \frac{1}{M_{\text{irr}}} \frac{dM_{\text{irr}}}{d(\ln t)} \right| = \left| \frac{d(\ln M_{\text{irr}})}{d(\ln t)} \right| = \frac{kT}{U_0}.
\end{equation}
Relaxation rate describes how rapidly the supercurrent decays following a logarithmic timescale. A larger relaxation rate signifies a faster loss of current-carrying capability, and in this linear-barrier approximation, measuring relaxation rate directly probes the pinning barrier height $U_0$. While high-temperature superconductors often exhibit nonlinear $U(J)$ relationships and collective creep effects that cause deviations from the simple $kT/U_0$ behavior, $S$ remains the standard metric for comparing flux creep dynamics across different temperature and magnetic field regimes practically. The magnetic relaxation measurement can help clarify the vortex state. In the elastic state, vortices can move easily, leading to a high relaxation rate. With the increase of the external magnetic field, the distance between the vortices decreases. Consequently, the repulsive forces between the vortices become stronger, leading to a decrease in vortex relaxation rate. With further increase in magnetic field, plastic motion of vortices takes over elastic one as the energy barrier for the plastic motion decreases with field, leading to an increase in the relaxation rate. The crossover between the elastic and plastic states is recognized as one of the origins of the peak in $J_{\rm c}$-$H$ curves for superconductors.

Figure \ref{fig:4}(a) shows the magnetic field dependence of the normalized relaxation rate $S$ at several temperatures for the pristine Hg1223 single crystal. The dip in $S(H)$ observed in Fig. \ref{fig:4}(a) is qualitatively consistent with the expected behavior described above. Comparing the relaxation data in Fig. \ref{fig:4}(a) with the $J_{\rm c}$ peaks, we note that the dip in $S(H)$ does not exactly coincide with the $J_{\rm c}$ peak position. This discrepancy is likely due to the low irreversibility field $H_{\text{irr}}$ of Hg1223, which limits the relaxation measurements at higher fields and temperatures as discussed later in the manuscript. Based on the above analysis, we construct a schematic phase diagram for the pristine Hg1223 single crystal as shown in Fig. \ref{fig:4}(b). The elastic-plastic crossover boundary is determined from the $J_{\rm c}$ peaks. The feature of the magnetic relaxation rate for pristine sample in Fig. \ref{fig:4}(a) will be discussed later with the irradiated samples. One more point about the peak effect in Hg1223 single crystals we should highlight is that in the previous study, a total of three $J_{\rm c}$-$H$ peaks have been reported in Hg1223 single crystals including the zero-field peak \cite{Ye2025Three}. A similar phenomenon of three peak effects has also been observed in NdBa$_2$Cu$_3$O$_{7-\delta}$ single crystal \cite{Koblischka1998BSCOOThree}. The missing peak in this study is the peak which originates from the order-disorder transition. The order-disorder transition induced peak effect can be easily affected by artificial defects. For example, in Ref. \cite{Koblischka1998BSCOOThree}, by oxygen annealing, Nd$_4$Ba$_2$Cu$_2$O$_{10}$ inclusions have been created in NdBa$_2$Cu$_3$O$_{7-\delta}$ single crystal, and in this sample, only two peaks are observed. Alternatively, disorder induced by proton irradiation may have broadened the order-disorder transition as observed in proton-irradiated NbSe$_2$ \cite{li2023PeakEffectsInduced}. Crystals used in this study have lower $T_{\rm c}$ compared to Ref. \cite{Ye2025Three}, indicating that the effect of natural defects could be stronger. In addition, as shown in Fig. \ref{fig:1}(f), the $J_{\rm c}$ data in the high-temperature and high-field region is quite noisy. This noise may also prevent us from clearly observing the third peak effect.

\begin{table*}[]
    \centering
    \caption{${\alpha}$ value in several superconductors after proton irradiation. All samples are at or near optimal doping.} \label{Table2}
    \begin{tabular}{c|c|c|c|c|c|c}
    \hline
    Sample & HgBa$_2$Ca$_2$Cu$_3$O$_{8+x}$ & YBa$_2$Cu$_3$O$_{8+x}$ & KCa$_2$Fe$_4$As$_4$F$_2$  & FeSe & Ba$_{0.6}$K$_{0.4}$Fe$_2$As$_2$ & Ba$_{0.6}$K$_{0.4}$Fe$_2$As$_2$\\
    \hline
    Particle     & 3 MeV Protons & 3 MeV Protons & 3 MeV Protons & 3 MeV Protons & 3 MeV Protons & 4 MeV Protons \\
    \hline
    Dose (/cm$^2$)    & $3\times10^{16}$ & $1\times10^{16}$ & $10\times10^{16}$ & $5\times10^{16}$ & $5.8\times10^{16}$ & $7\times10^{16}$\\
    \hline
    {$\alpha$} exponent     & 0.98 & 0.6 \cite{civale1991vortex} & 0.5 \cite{pyon2021critical} & 0.3 \cite{Sun2015FeSealpha}  & 0.3 \cite{taen2015CriticalCurrentDensity} & 0.47 \cite{Kihlstrom2013Bak1224MeV}\\
    \hline
    \end{tabular}
    \label{tab:placeholder}
\end{table*}

\begin{figure*}[]
\centering
\hspace{-1mm}\includegraphics[scale=0.8]{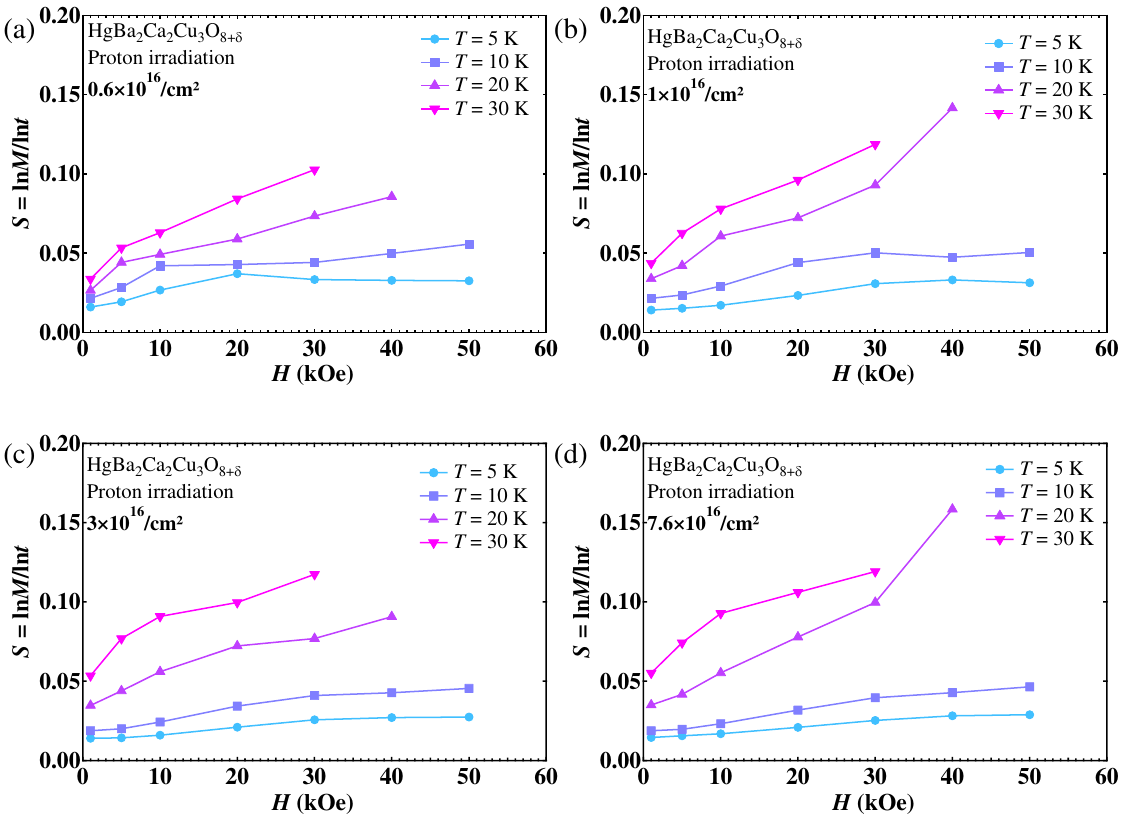}
\caption{\label{fig:5}External magnetic field dependence of magnetic relaxation rate for Hg1223 single crystals after 3 MeV proton irradiation with dose of (a) $0.6\times10^{16}$/cm$^2$, (b) $1\times10^{16}$/cm$^2$, (c) $3\times10^{16}$/cm$^2$, and (d) $7.6\times10^{16}$/cm$^2$. }
\end{figure*}

For the $J_{\rm c}$-$H$ curves of the irradiated samples, the $\alpha$ value becomes larger. As indicated by the dashed lines in Figs. \ref{fig:1}(g)-(i). The slope for crystals irradiated with doses of $0.6\times10^{16}$/cm$^2$, $1\times10^{16}$/cm$^2$, and $3\times10^{16}$/cm$^2$ at $T = 15$ K are approximately 0.87, 0.95, and 0.98, respectively. These values are relatively high compared to other superconductors after proton irradiation. For example, in Table \ref{Table2}, several ${\alpha}$ values for different superconductors are summarized, where the ${\alpha}$ value for Hg1223 shows the largest value among them. One possible reason for the large exponent in Hg1223 may be due to the large magnetic relaxation induced by low $H_{\rm irr}$. The magnetic relaxation is expected to diverge at $H_{\rm irr}$. As reported in Ref. \cite{Bras1996anisotropic}, the anisotropy of Hg1223 is 52. $H_{\rm irr}$ decreases significantly with increasing temperature, causing the relaxation rate to increase sharply with the magnetic field in the high-temperature region. Conversely, the behavior of magnetic relaxation also affects $J_{\rm c}$-$H$ characteristics, leading to a monotonic decrease in $J_{\rm c}$ with increasing magnetic field, as shown in Figs.\ref{fig:1}(f)-(j). The absence of a simple power-law dependence of $J_{\rm c}$ on magnetic field in the high-temperature region can be attributed to this reason. Furthermore, when $H_{\rm irr}$ is low, magnetic relaxation measurements at the same time scale are not the same for different magnetic fields, as discussed in Ref. \cite{pyon2021critical}. The low $H_{\rm irr}$-induced large relaxation at high magnetic fields and high temperatures causes the measured $J_{\rm c}$ to be less than the real $J_{\rm c}$, thereby resulting in a larger ${\alpha}$ value.

Figures \ref{fig:5}(a)-(d) show the normalized magnetic relaxation rate $S$ of irradiated Hg1223 single crystals as functions of magnetic field at different temperatures. The magnetic field dependence of $S$ after irradiation is significantly different from that before irradiation. In Fig. \ref{fig:4}(a), characteristic peaks in $S$-$H$ are formed at low fields. Such a feature for pristine Hg1223 single crystals is similar to previous studies on YBa$_2$Cu$_3$O$_{7-\delta}$ \cite{yeshurun1989MagneticPropertiesYBaCuO} and Hg1201 \cite{eley2020VortexPhasesGlassy}. On the other hand, the $S$ value of irradiated samples increases monotonically with magnetic field as shown in Figs. \ref{fig:5}(a)-(d). One possible explanation for the lack of low-field peaks and the different temperature dependence of $S$ after irradiation in Hg1223 is that the introduction of artificial defects through irradiation makes the formation of the single vortex pinning difficult. The absence of the single vortex regime could lead to the monotonic increase of $S$ with temperature even at low fields. Another feature of $S$-$H$ for irradiated samples is that the normalized relaxation rate increases strongly with magnetic field above 20 K. This behavior, once again, can be attributed to low $H_{\rm irr}$ in Hg1223, since the lower $H_{\rm irr}$ will cause the relaxation to diverge earlier. From the $J_{\rm c}$ contour map in Figs.\ref{fig:1}(k)-(o), it seems that $H_{\rm irr}$ is slightly expanded after irradiation. However, at the same time, $T_{\rm c}$ is suppressed. This can also be attributed to the monotonic increase of the relaxation rate for all irradiated crystals.

It is important to note that in the magnetic relaxation study of cuprate superconductors, a well-known phenomenon is the observation of a plateau (glassy state) in the temperature dependence of the normalized relaxation rate, with relaxation rate values in the plateau ranging from 0.02 to 0.04 \cite{malozemoff1990UniversalityCurrentDecay}. This phenomenon has been observed in YBCO single crystals \cite{yeshurun1996magnetic}. In the present study, this phenomenon is not pronounced for neither pristine nor irradiated Hg1223 single crystals. From the magnetic field dependence of $S$ in Fig. \ref{fig:4} and Fig. \ref{fig:5}, it can be seen that most of the magnetic relaxation rates already exceed the threshold of 0.04 even at 30 K. One possible reason for the absence of the plateau in Hg1223 single crystals could be attributed to the relatively low $H_{\rm irr}$. The $H_{\rm irr}$ of pristine Hg1223 single crystals is approximately 12 kOe at $T = 77$ K, whereas it is around 80 kOe for YBCO at the same temperature \cite{civale1991vortex}. 

\begin{figure*}[]
\centering
\hspace{-1.5mm}\includegraphics[scale=0.65]{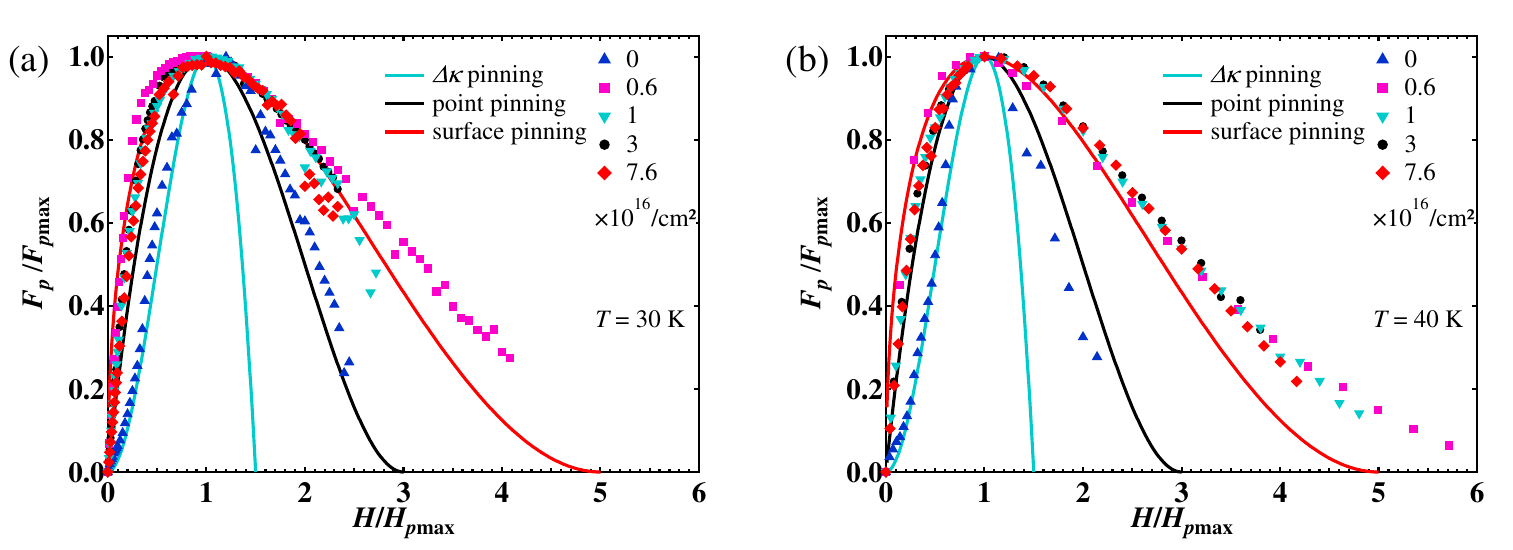}
\caption{\label{fig:6}The reduced magnetic field dependence of reduced pinning force density for Hg1223 single crystals before and after 3 MeV proton irradiation at (a) $T = 30$ K and (b) $T = 40$ K.}
\end{figure*}

In order to better understand the pinning mechanisms of Hg1223 before and after 3 MeV proton irradiation, the extended Dew-Hughes method is used for analysis. Pinning mechanisms of vortices in superconductors can be categorized into several types \cite{dew1974flux}. In the standard Dew-Hughes model, the reduced pinning force density is expressed as $f_p \propto h^p(1-h)^q$, where $p$ and $q$ are parameters characterizing different pinning mechanisms and $h = H/H_{\text{c2}}$. For conventional superconductors, $H_{\text{c2}}$ is used directly. For high-temperature superconductors with large thermal fluctuations, $H_{\text{c2}}$ is often replaced by the irreversibility field $H_{\text{irr}}$. The function $f_p$ takes a maximum at $h_{\text{max}} = p/(p+q)$. In the case of surface-like pinning, $h_{\text{max}} = 0.2$. For point-like pinning, $h_{\text{max}} = 0.33$. However, the irreversibility field $H_{\text{irr}}$ at 40 K for Hg1223 already exceeds the maximum field of our MPMS (50 kOe) as can be seen from the contour maps in Fig. 1. For irradiated samples, $H_{\text{irr}}$ is even larger. Therefore, an accurate determination of $H_{\text{irr}}$ is not possible in this field range. To overcome this difficulty, we adopt the extended Dew-Hughes method \cite{higuchi1999comparative}, in which the reduced field is redefined as $h = H/H_{p\text{max}}$, where $H_{p\text{max}}$ is the field at which the pinning force density reaches its maximum $F_{p\text{max}}$. In this case, the relation for the representative three pinning mechanisms are
\begin{equation}
f_p = 3h^2(1-2h/3),    \quad(\Delta\kappa \text{ pinning}),
\end{equation} 
\begin{equation}
f_p = 9h(1-h/3)^2/4,    \quad(\text{point pinning}),
\end{equation}
\begin{equation}
f_p = 25h^{1/2}(1-h/5)^2/16,    \quad(\text{surface pinning}).
\end{equation}

Figure \ref{fig:6} show the reduced pinning force density as a function of the reduced magnetic field $h$ for Hg1223 irradiated by 3 MeV protons at different doses measured at $T = 30$ K and 40 K. These two temperatures are chosen because at lower temperatures the pinning force does not exhibit a peak within the available field range, while at higher temperatures thermal activation becomes more pronounced. For the pristine sample, the pinning mechanism can be well fitted by $\Delta\kappa$ pinning at low magnetic fields, while at high magnetic fields, it falls between $\Delta\kappa$ pinning and point pinning. This result is similar to the previous study on NdBa$_2$Cu$_3$O$_y$ \cite{higuchi1999comparative}. The synthesis process of the crystals inherently leads to the formation of defects such as dislocation and Frenkel pairs, resulting in a combination of multiple pinning mechanisms. After irradiation by 3 MeV protons, a significant shift in the pinning mechanism is observed. At low magnetic fields, the pinning mechanism can be well fitted by point pinning, which is reasonable since proton irradiation primarily introduces a large number of point defects that serve as effective pinning centers. This explains the observed enhancement of $J_{\rm c}$ at low magnetic fields in Fig. \ref{fig:1}. In the high magnetic field regimes, a pronounced change in mechanisms was indeed observed before and after 3 MeV proton irradiation. However, as discussed above, the magnetic relaxation at high magnetic fields is significantly large, which strongly affects the pinning force. Therefore, the true pinning mechanisms at high magnetic fields may be obscured by the effects of magnetic relaxation.  
\\
\section{Summary}
The pronounced enhancement of $J_{\rm c}$ in Hg1223 single crystals has been achieved through 3 MeV proton irradiation. The self-field $J_{\rm c}$ at 77 K for all the irradiated crystals has improved to exceed 0.1 MA/cm$^2$. The large value of exponent $\alpha$ in the power-law dependence of $J_{\rm c} \propto H^{-\alpha}$ is observed after 3 MeV proton irradiation, which can be attributed to the low magnetic irreversibility field resulting from the large anisotropy in Hg1223 single crystals. Analysis of the reduced pinning force density fitted using the extended Dew-Hughes model reveals a clear change in pinning mechanisms after 3 MeV proton irradiation. 
\\
\ack{This work was supported by Jiangsu Funding Program for Excellent Postdoctoral Talent under Grant Number 2024ZB282 and 2025ZB141, the CAS Superconducting Research Project under Grant No. [SCZX-0101],  the National Natural Science Foundation of China (Grant No. 12374135, Grant No. 12374136, and Grant No. 12304193), Key Program of National Natural Science Foundation of China (Grant No. U24A2068), the National Basic Research Program of China (Grant No. 2025YFA1411501), the Southeast University Interdisciplinary Research Program for Young Scholars, the Fundamental Research Funds for the Central Universities (Grant No. 2242025F10008), the authors thank the Center for Fundamental and Interdisciplinary Sciences of Southeast University for the support in transport measurements. This work is partly supported by a Grant-in-Aid from MEXT in Japan and is part of the Research Project with Heavy Ions at NIRS-HIMAC. W. Li, R. Guo and X. Zhou contributed equally to this paper.}

\balance
\printorcids
\providecommand{\noopsort}[1]{}\providecommand{\singleletter}[1]{#1}%
\providecommand{\newblock}{}

\end{document}